\documentclass[a4paper,11pt,sort&compress,number,5p]{elsarticle}
\usepackage[utf8]{inputenc}
\usepackage{lineno,hyperref}
\usepackage{pifont}
\usepackage{natbib}
\usepackage{fleqn}
\usepackage{graphicx}
\usepackage{txfonts}
\usepackage{multicol}
\usepackage{multirow}
\usepackage{caption}
\usepackage{tabularx}
\usepackage{floatflt}
\usepackage{xcolor}
\usepackage{booktabs}
\usepackage{dcolumn}
\modulolinenumbers[5]
{}
\journal{Materials Today Communications}

\begin{document}

\begin{frontmatter}

\title{Ab initio study of the half-metallic full-Heusler compounds Co$ _2 $ZAl [Z\,=\,Sc, Ti, V, Cr, Mn, Fe]; the role of electronic  correlations}
\author[1]{S.~Nepal}
\ead{sashinepal36@gmail.com}
\author[1]{R.~Dhakal\corref{cor1}}
\ead{ramesh.dhakal91@gmail.com}
\author[2]{I. Galanakis}
\ead{galanakis@upatras.gr}

\cortext[cor1]{Corresponding author}

\address[1]{Central Department of Physics, Tribhuvan University, Kathmandu, Nepal}
\address[2]{Department of Materials Science, School of Natural Sciences, University of Patras, GR-26504 Patras, Greece}
\begin{abstract}
We study the structural, electronic, and magnetic properties of Co$_2$ZAl compounds employing a pseudopotential electronic band structure method. The stability of the compounds is established through formation and cohesive energy calculations. The effect of the lattice parameter variation on the electronic and magnetic properties of the compounds is investigated and meticulous explanation is provided for the observed behavior. The variation of the individual spin magnetic moments and the stability of the total spin magnetic moment during the expansion and contraction of the lattice parameter is observed and an attempt is made to understand the obtained behavior. Finally, we implement DFT$+U$ to examine its consequences on the electronic and magnetic properties of the Co$_2$ZAl compounds. We find that the use of DFT$+U$ is not justified for these compounds and in some cases like Co$_2$MnAl it produces unrealistic properties. Exception is  Co$_2$FeAl where the desired half-metallicity is restored  after the inclusion of on-site correlations. We explain why the on-site correlations might be important for Co$_2$FeAl by comparing it with other Heusler alloys where the correlation was found to be meaningful to explain the observed magnetic moments.                 	

\end{abstract}

\begin{keyword}
Cobalt based full Heusler alloy\sep DFT$ +U $ \sep Half-Metallicity, ab initio
\end{keyword}

\end{frontmatter}

\section{Introduction}\label{sec1}

Heusler alloys, several members of which are well-known for their special properties like high Curie temperature, half metallicity, and exceptional tunability, is a large family of ternary intermetallic compounds with diverse magnetic phenomena.  Particularly, half-metallic Heuslers are very interesting because of the high spin-polarized current near the Fermi level and are expected to increase the efficiency of spintronic devices\cite{spintronics,Hirohata,SpintronicsClaudia}. The different nature of the two spin channels in half-metals, one insulating and the other conducting, is a novel property which can be exploited to enhance the performance of information storage and spin injecting devices.   Among the diverse half-metallic Heusler alloys, special attention is given to the Co-based compounds since they have been widely explored experimentally. In addition their high Curie temperatures makes them promising candidates for room-temperature applications. Many of them have been investigated for giant magnetoresistive devices, devices based on the anisotropic magnetoresistance effect, and magnetic tunnel junctions\cite{HMAlloys-Galanakis-lec, HeuslerPropandGrowth}.    
                        
\par In 1983, de Groot, in his groundbreaking paper\cite{deGroot}, found that the semi-Heusler alloy NiMnSb shows an asymmetric band structure for different spin polarization of the electrons. Although the Ni atoms have a largely neutral character for both spins, they noticed  a different characteristic behavior for the Mn \textit{d} electrons leading to 100\% spin polarization near the Fermi level. The same year, K\"{u}bler \textit{et al.} explained how the spin magnetic moments of different atoms in Heusler alloys are formed and coupled with each other\cite{kubler}, giving the detailed picture for different types of magnetic ordering. They also noticed that the minority-spin state density, for cobalt-based full Heusler compound under study, nearly vanishes at Fermi level and underscored the fact that it could lead to anomalous transport properties. Since then, the research in Heusler compounds has been largely intensified and special attention is given to Co-based Heusler compounds.

\par Almost two decades after the initial synthesis of Co$ _2 $YZ compounds by Ziebeck  and Webster \cite {webster,Ziebeck},  Ishida \textit{et al.} extensively studied the Co$ _2 $MnZ compounds\cite{Ishida} and found that Co$ _2 $MnSi  and Co$ _2 $MnGe  are half-metallic in nature. Next year, Carbonari \textit{et al.} studied Co$ _2 $YZ compounds and measured the magnetic moment of Cobalt sites by using hyperfine field measurement \cite{Carbonari}.  In 2002, Galanakis \textit{et al.} described  the origin of the half-metallic behavior  in full Heusler compounds \cite{Galanakis2002} and presented the possible hybridization scheme of the orbitals taking  Co$ _2 $MnGe as example. In the same article, they established the Slater Pauling rule for full Heusler compounds, which is of great importance since it relates the total spin magnetic moment of a particular compound and the total number of valence electrons, and consequently can be used to predict the half-metallic nature of the compounds. They later extended the study of the Slater Pauling behavior for other structures of Heusler compounds\cite{Galanakisquaternary,Galanakisinverse}.

\par A detailed study of a large number of cobalt-based full Heusler alloys was made by Kandpal and collaborators in 2006. In their article\cite{Kandpalmainpaper},  half-metallic ferromagnets of different types were classified and meticulously analyzed. They showed that the inclusion of electronic correlations for Heusler alloys does not have a significant effect on the half-metallic gap and their inclusion is justified only for Co$ _2 $FeSi which has the maximum number of valence electrons per unit cell required for  half-metallicity to be present \cite{KandpalcompareMnFe}. Galanakis \textit{et.al} have also studied the effect of electronic correlations, accounted through a Hubbard $U$ parameter, on a large number of half-metallic magnets and showed that inclusion of $U $ produces a more or less similar effect on Heusler alloys\cite{dftueffect}. It is important to note that in their study, in some cases, they have found that the inclusion of $ U $ produced unrealistic results. 

\par Since most of the studies mentioned above concern Co-based Heusler compounds where the metalloid atom is Si, in the present study we focus on the Co-based compounds where the metalloid atom is Al. All compounds Co$_2$ZAl with Z ranging from Ti to Fe have been experimentally grown \cite{Kandpalmainpaper,BUSCHOW19831,BUSCHOW198190,cta}. For reasons of completeness we consider compounds of the Co$_2$ZAl  formula where  Z can be one of the transition metal atoms from Sc to Fe. For our calculations, we adopt a pseudopotential ab-initio electronic band structure method to study the structural, electronic, magnetic and gap related properties of these compounds. Since in most cases these compounds are grown in the form of thin films, we expand our study to the case of both lattice contraction and expansion. Finally, we pay special attention to the role of electronic correlations on the electronic and magnetic properties of these compounds. The article is organized as follows. In section \ref{sec2} we present the computation details of our calculations. In section \ref{sec3} we present our results and in section \ref{sec4} we summarize and present our conclusions.

\section{Computational details}\label{sec2}

\par
\begin{figure}[h]
	\centering
	\includegraphics[width=0.9\linewidth]{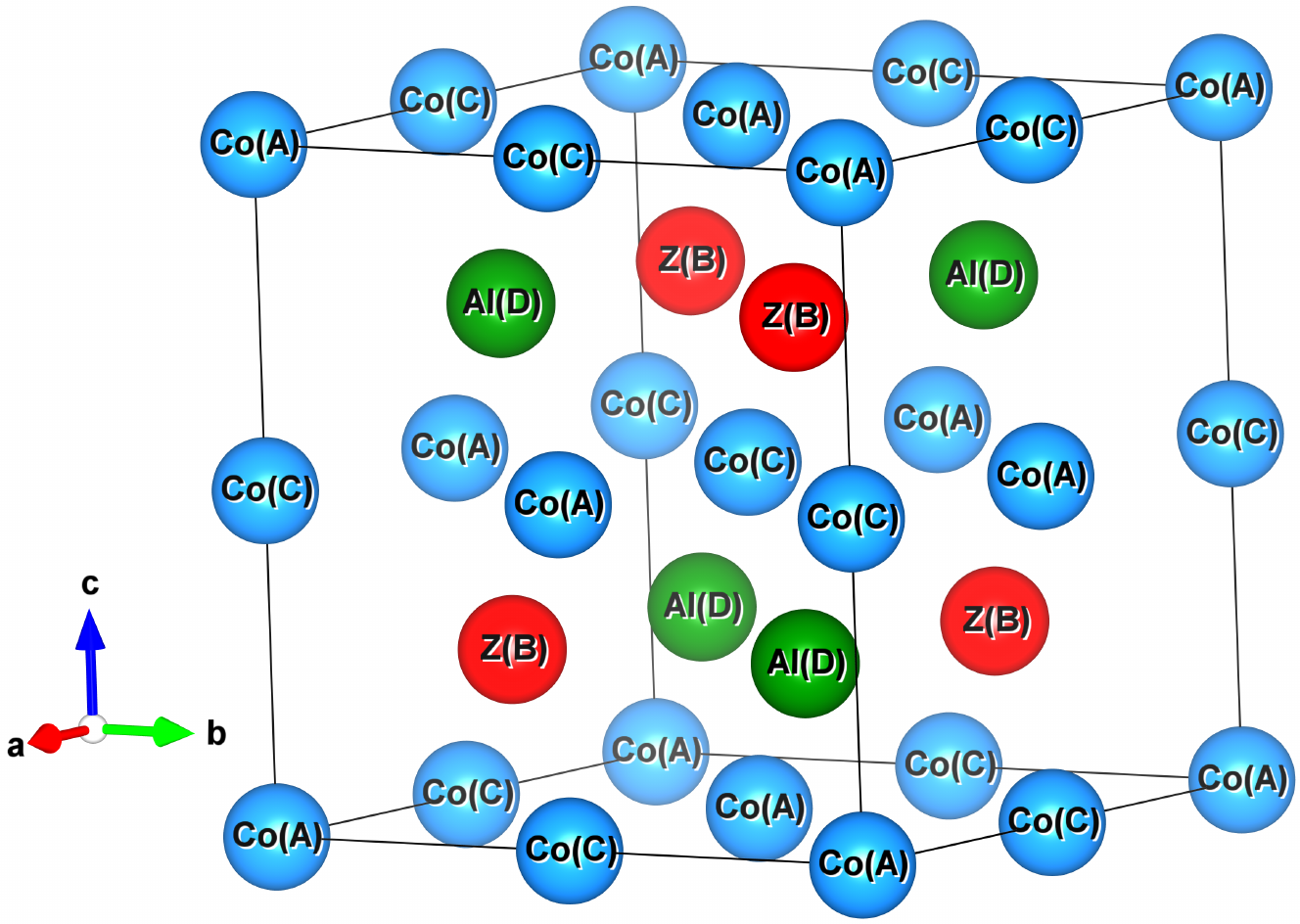}
	\caption{L2$_1$ crystal structure of Co$ _2 $ZAl [Z\,=\,Sc, V, Ti, Cr, Mn, and Fe] full-Heusler compounds. Here, the letter inside bracket in figure represents the Wyckoff position mentioned above.}
	\label{fig:cza}
\end{figure}

 X$ _2 $YZ type Heusler compounds crystallize in the cubic L2$ _1 $ and XA structures which both consist of four interpenetrating fcc lattices with Wyckoff positions: A(0, 0, 0), B(0.25, 0.25, 0.25), C(0.5, 0.5, 0.5) and D(0.75, 0.75, 0.75). Here, X and Y represent transition metals such as Mn, Fe, Co, etc. and Z represents main group elements such as Al, Ga, Si, etc. Depending upon the number of valence electrons of the X and Y elements, a Heusler alloy can crystallize either in the  L2$ _1 $-type or XA-type structures. If X has more (less) valence electrons than Y then the compound prefers the L2$ _1 $-type (XA-type) structure in which the X elements occupy the  positions A and C(B) mentioned above. The so-called semi(or half)-Heusler compounds, which have the chemical formula XYZ, prefer the Cl$ _b $-type cubic structure where the C site remains empty. In the present study, the full-Heusler compounds Co$ _2 $ZAl [Z\,=\,Sc, V, Ti, Cr, Mn, Fe] crystallize in the L2$_1$ structure shown in figure \ref{fig:cza}.

\par The electronic and magnetic properties of the materials under study are computed using the plane-wave pseudopotential based method as implemented in the Quantum Espresso Package [v 6.5]\cite{QE-2009,QE-2017} in conjunction with the Density Functional Theory (DFT). For  Aluminum, Scandium, Manganese, and Iron atoms the Projector-Augmented Waves (PAW) basis set is used and for Titanium, Vanadium, Chromium, and Cobalt atoms ultra-soft pseudopotentials are exploited. The pseudopotential files for our calculation are obtained from \hyperlink{https://www.materialscloud.org/}{www.materialscloud.org}\cite{psedopotentialcite1,psedopotentialcite2}. We estimate the exchange-correlation functional using the Perdew-Burke-Ernzerhof parametrization corrected for solids of the generalized gradient approximation  (PBEsol GGA)\cite{pbesol1,pbesol2,pbesol3}. The cutoff energy for the plane-waves is chosen to be 175 Ry and  for the  charge density ten times larger (1750 Ry). A Monkhorst-Pack grid of  8\,x\,8\,x\,8 size is considered for the k-points sampling in the electronic structure calculations. For the calculation of the density of states (DOS), a denser Monkhorst-pack grid of 12\,x\,12\,x\,12 size is considered and the linear tetrahedral method for the integration\cite{lineartetrahedra} is used. Structure optimization is performed using the total energy minimization method. The convergence thresholds for the self-consistent cycle and the  total energy are set to $ 10^{-8} $ and $10^{-5}$ Ry, respectively.

\par To study the on-site electronic correlations effect on the electronic and magnetic properties of the compounds, we perform DFT$+U$ calculation where $U$ is the Hubbard parameter. In our calculations, we use the rotationally invariant DFT$+U $ scheme of Liechtenstein \textit{et al}. in which the Hubbard $ U $ and exchange $ J $ parameters are treated separately\cite{Liechtenstein}. Since the Coulomb repulsion energy among correlated states is already included in GGA calculations, in the DFT+$U $ scheme, the introduction of Hubbard $ U $ elicits the need for a double counting correction term in the energy functional. So, we need to remove the Coulomb energy part from the DFT functional, accounted by GGA for correlated states. These are two functionals--the around mean-field (AMF) functional and the fully localized limit (FLL) functional--which are common choices for the double-counting term.  In Quantum Espresso package the fully localized limit (FLL) functional is implemented.

\par The value of the $U$ and $J$ parameters depend upon the system under study and have to be determined for each system separately,  which is a very tedious and difficult task. Also, the experimental determination of parameters is very difficult and only scarce data exists. Thus, one has to rely on ab-initio calculations of these parameters. In literature \cite{dFTparams}, Şaşıoğlu \textit{et. al.} have calculated the Hubbard $U$ and exchange $J$ parameters for several full-Heusler compounds implementing the constrained Random Phase Approximation (cRPA) and we have used these parameters in our DFT$+ U $ calculations. The values of Hubbard $U$ and exchange $J$ parameters in eV for transition metals are tabulated in table \ref{tab:hubbardU}. The calculated properties of Co$_2$ZAl compounds using the GGA$+U$ functional are tabulated in table \ref{tab:dftuprop} and will be discussed later. Finally, one should note  that although within the DFT$+U$ scheme we converge the total energy, the parameter nature of $U$ in LDA$ +U $ or GGA$ +U $ \, results in an artificial opening of the gaps in semiconductors while in magnets the exchange splitting of the occupied majority-spin and unoccupied minority-spin increases. Since the LDA$ +U $ and GGA$ +U $ functionals are not variational with respect to $ U $ itself, one can not safely conclude which of the properties obtained, with or without $ U $, are the real ones and whether the inclusion or not of $U$ is  adequate to describe the material.

\begin{table}[hp]
	\caption{Hubbard $ U $ and exchange $ J $ parameters\cite{dFTparams} for 3d transition metals in compounds under study taken from reference \cite{dFTparams}.}
	\label{tab:hubbardU}
	\begin{tabular}{ c c c c }
		\toprule
		Compounds                                         &                            & \multicolumn{1}{l }{$U$ (eV)} & \multicolumn{1}{c }{$J$ (eV)} \\ \toprule
		\multirow{2}{*}{Co$_2$CrAl}                       & \multicolumn{1}{c }{Co-3d} & 3.74                          & 0.72                          \\ 
		& Cr-3d                      & 2.82                          & 0.54                          \\ \hline
		\multicolumn{1}{ c }{\multirow{2}{*}{Co$_2$MnAl}} & Co-3d                      & 3.40                          & 0.70                          \\ 
		\multicolumn{1}{ c }{}                            & Mn-3d                      & 3.23                          & 0.58                          \\ \hline
		\multirow{2}{*}{Co$_2$FeAl}                       & Co-3d                      & 3.00                          & 0.66                          \\ 
		& Fe-3d                      & 3.43                          & 0.66                          \\ \bottomrule
	\end{tabular}
\end{table}

\section{Result and discussions}\label{sec3}

\subsection{Structural optimization and stability}

\par
\begin{figure}[hpt]
	\centering
	\includegraphics[width=1\linewidth]{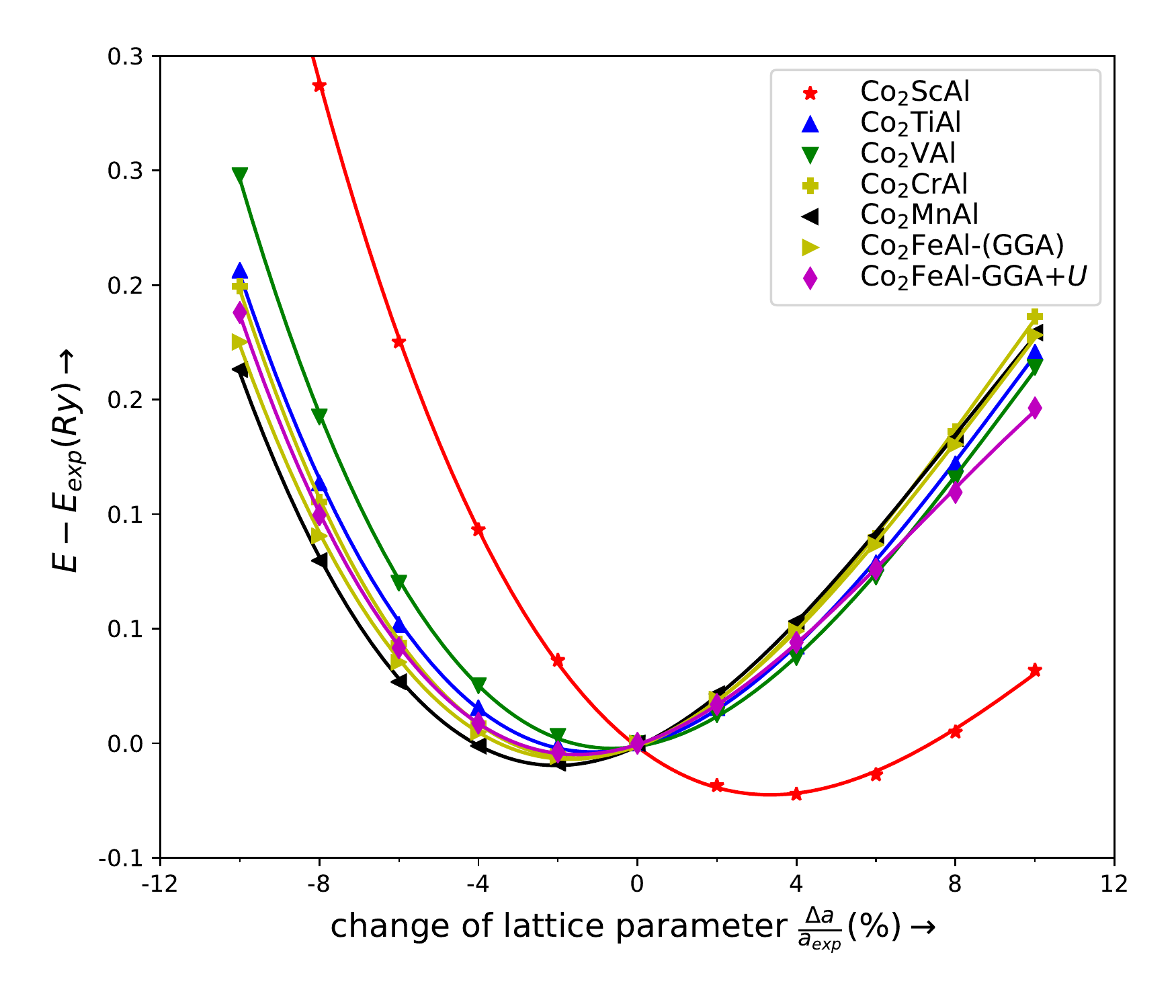}
	\caption{Structural optimization for the compounds Co$_2$ZAl [Z = Sc, Ti, V, Cr, Mn and Fe]. We plot the total energy (E-E$_{exp}$) as a function of the lattice parameters. Here, E$_{exp}$ represents the total energy of a compound at the experimental lattice parameter except for Co$_2$ScAl. The x-axis is the change in the lattice parameter(\%) with respect to the experimental value. }
	\label{fig:02}
\end{figure}

 The structural optimization of the compounds Co$_2$ZAl [Z = Sc, Ti, V, Cr, Mn, Fe] is performed using the total energy minimization method. We obtain the total energy per formula unit of compounds for different lattice parameters with a step of  $2\%$ and adopting in all cases a cubic lattice. For all compounds we start from the experimental lattice constant (\cite{Kandpalmainpaper,BUSCHOW19831,BUSCHOW198190,cta}) and  vary up to a maximum(minimum) of $\pm 10\%$. Exception is Co$_2$ScAl which  is the only compound which has not  been grown experimentally. We have chosen its starting lattice constant to be 5.722 $\AA{}$, identical to the experimental lattice constant of Co$_2$VAl. For Co$_2$FeAl, also a GGA$+U $ calculation is carried out. The results of the structure optimization are shown in figure-\ref{fig:02}. The values of the optimized parameters, as well as other physical quantities computed, are tabulated in table \ref{tab:table3} and will be discussed in detail in later sections.

\par The  stability of the compounds structure is demonstrated by the calculation of the formation and cohesive energies. The formation and cohesive energies for Co$_2$ZAl [Z = Sc, Ti, V, Cr, Mn, \& Fe] can be expressed as
\begin{equation}
\label{eq1}
E_{Form}^{Co_2ZAl}=E_{tot}-(2E_{Co}^{Bulk}+E_{Z}^{Bulk} + E_{Al}^{Bulk})
\end{equation}
\begin{equation}
\label{eq2}
E_{Coh}^{Co_2ZAl}=E_{tot}-(2E_{Co}+E_{Z} + E_{Al})
\end{equation}

\par Here, $E_{Form}^{Co_2ZAl}$ and $E_{Coh}^{Co_2ZAl}$ stand for the formation and cohesive energies of the compounds respectively and the total energy per formula unit is denoted by $E_{tot}$. In expression-\ref{eq1}, $E_{Co}^{Bulk}$, $E_{Z}^{Bulk}$, and $E_{Al}^{Bulk}$ denote the total energy per atom for the  bulk of the element and in expression-\ref{eq2} $E_{Co}$, $E_{Z}$, and $E_{Al}$ denote the total energy of the corresponding single isolated atom. From table \ref{tab:tablcoh}, we can deduce that the cohesive energy for all the compounds are negative which establishes the structural stability of our compounds. We have also calculated the cohesive energy and the formation energy using the optimized parameters (not presented here) from which we reach the same conclusion.

\begin{table}[hp]
	\caption{Formation energy and cohesive energy for compounds under study calculated using the experimental lattice parameter.}
	\label{tab:tablcoh}
	\resizebox{0.48\textwidth}{!}{%
		\begin{tabular}{@{} l c c @{}}
			\toprule
			Alloy      & E$_{form}$(eV/atom) & E$_{coh}$(eV/atom) \\ \toprule
			Co$_2$ScAl & -0.523              & -5.589             \\ 
			Co$_2$TiAl & -0.627              & -5.946             \\ 
			Co$_2$VAl  & -0.421              & -5.814             \\ 
			Co$_2$CrAl & -0.209              & -5.279             \\ 
			Co$_2$MnAl & -0.314              & -5.333             \\ 
			Co$_2$FeAl (GGA ) & -0.340              & -4.541             \\
			Co$_2$FeAl (GGA$ +U $) & -0.494              & -4.561             \\ \bottomrule
		\end{tabular}%
	}
\end{table}

\subsection{Electronic and magnetic properties}

\begin{figure}[h]
	\centering
	\includegraphics[width=1\linewidth,height=0.5\textheight]{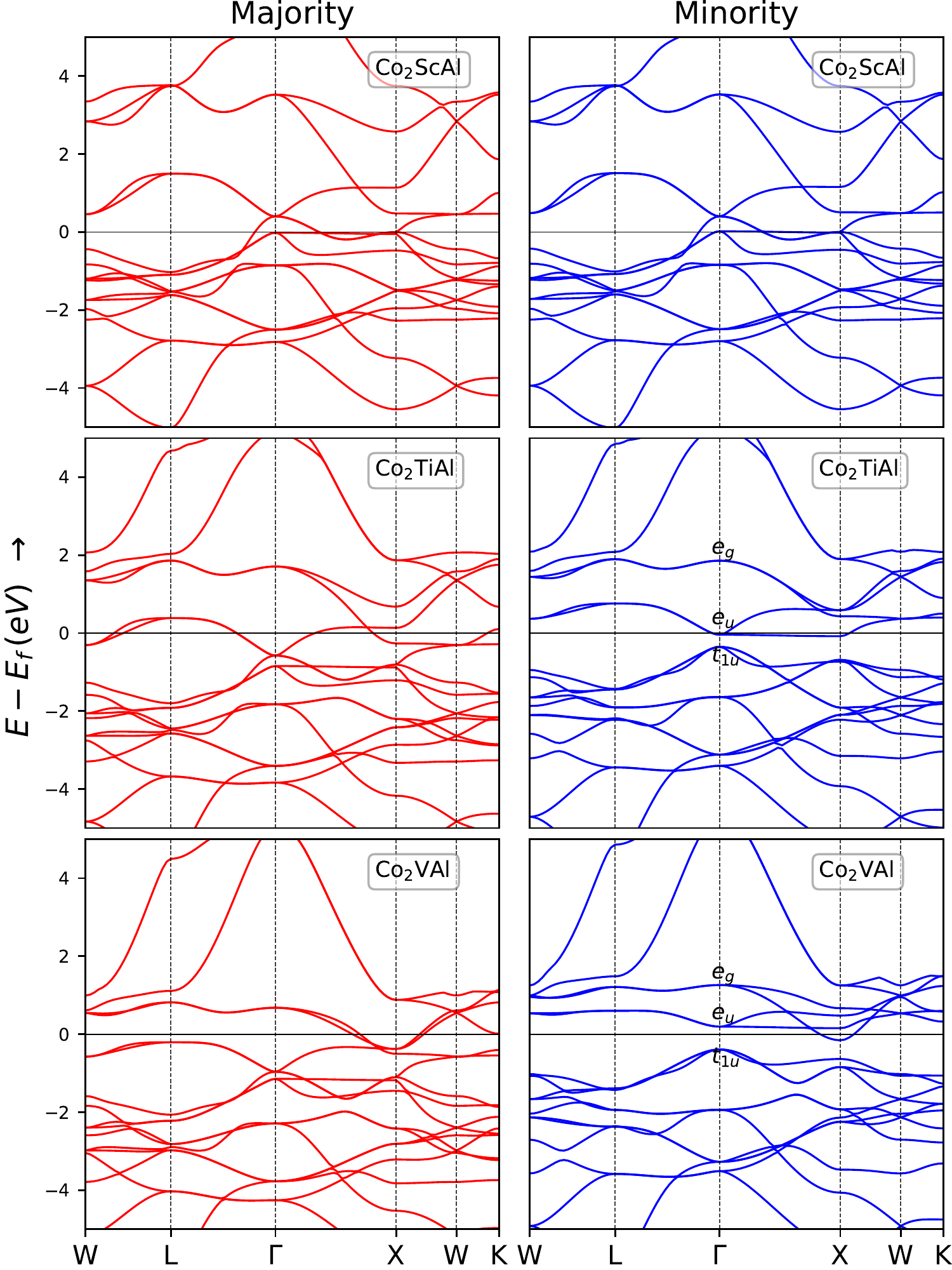}
	\caption{Calculated electronic structure of compounds Co$_2$ScAl, Co$_2$TiAl and Co$_2$VAl using optimized lattice parameter. Electronic
		structure was obtained from GGA calculations.}
	\label{fig:band1}
\end{figure}

\begin{figure}[h]
	\centering
	\includegraphics[width=1\linewidth,height=0.5\textheight]{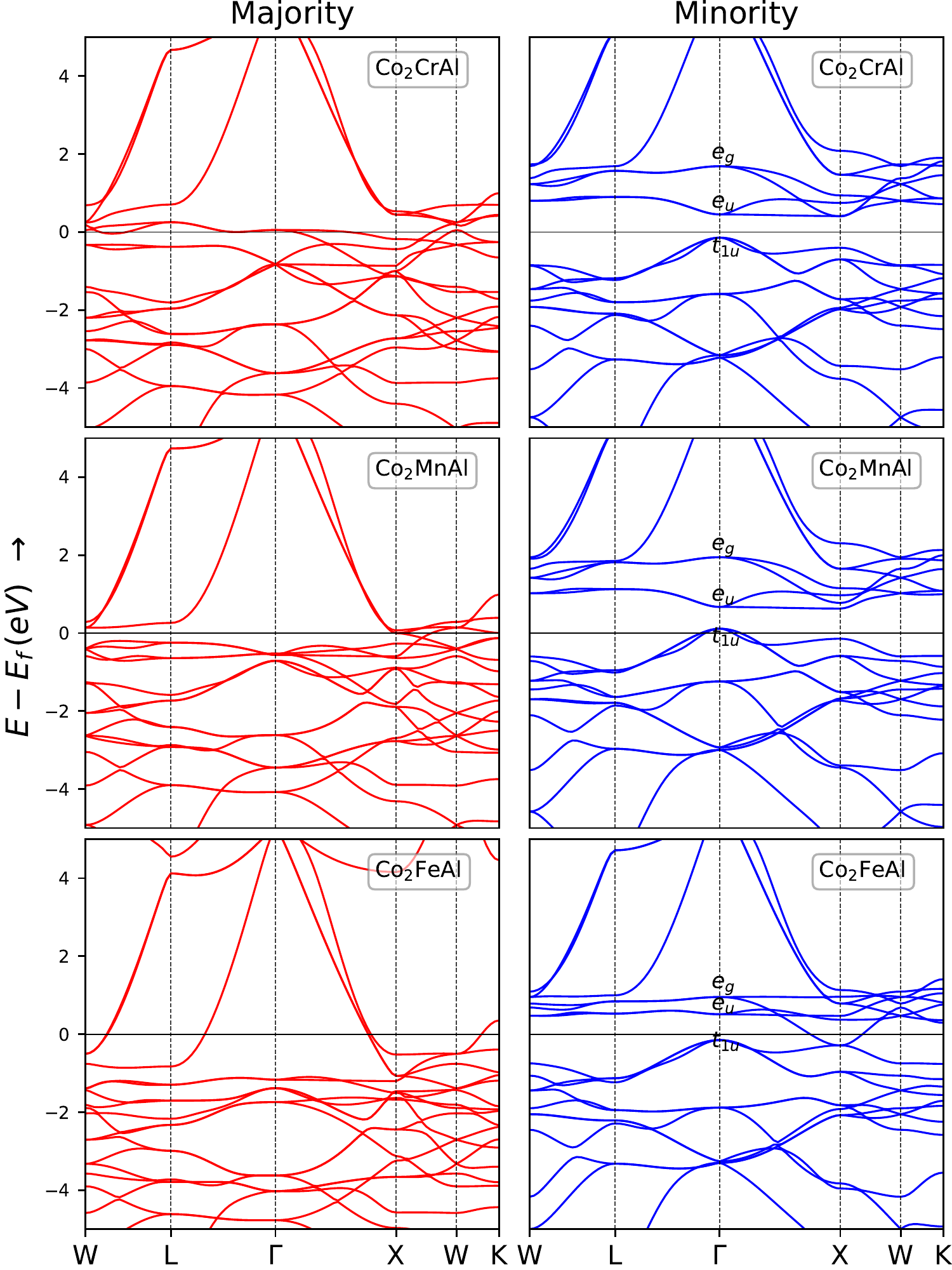}
	\caption{Calculated electronic band structure of the compounds Co$_2$CrAl, Co$_2$MnAl and Co$_2$FeAl using optimized lattice parameter. Result is obtained using GGA calculation.}
	\label{fig:band2}
\end{figure}

In figure \ref{fig:band1} and \ref{fig:band2} we present the majority and minority spin band structures of the Co$_2$ZAl compounds under study. The spin resolved band structure in each case is calculated using the optimized lattice parameter. The nature of the band structure of our calculations is in good agreement with the previously calculated band structures\cite{Co2CrAl,Co2ScAl,Co2VAl,Co2MnAl}.

\par The band structure of these compounds is well understood, thanks to the hybridization scheme provided by Galanakis \textit{et al.}\cite{Galanakis2002}, and  a meticulous description of the band structure is possible. In full Heusler alloys, the general idea is to understand that the Cobalt atoms at the two different sites, which are arranged  in octahedral symmetry in a crystal, hybridize first and then interact with the $d$ orbitals of Z atoms of the compound to form a complicated band structure around the Fermi level. The hybridization of different states of these atoms generates various bonding, non-bonding  and anti-bonding states which transform through different representations. In our explanation of the band structure, we are going to use this hybridization scheme. In our previous work, we have used a similar scheme for the  inverse Heusler alloys\cite{Galanakisinverse} to describe the band structure of Mn$_2$CoAl and Mn$_2$CoGa  compounds\cite{ourarticle}.

\par The occupied part of the minority spin states can be divided into different regions. The low lying \textit{s}-states of Al atoms, which is the same for both spin directions, is not shown in the figure. Above the \textit{s}-states lie the 3 \textit{p} states, which are formed as the result of hybridization of \textit{p} states of Co and Z atoms with the \textit{p} states of Al atoms. The states around the Fermi level are due to the strong hybridization between Co-Co and Co-Z atoms. Despite the detailed differences around the Fermi level and relative energy position of bands, the characteristics of the bands described above are valid for all the compounds under study. For the majority spin band structure, the explanation still holds and the only difference is that as we move from Sc to Fe an extra state is added due to the extra valence electron.

\par It is clear from the figure that except Co$_2$CrAl, all the other compounds slightly  deviate from the ideal half-metallic behavior. In the case of Co$_2$CrAl, one can see strong metallic behavior for the majority spin channel whereas there is a clear bandgap for the minority spin channel and the Fermi level lies between the valence and the conduction band. The gap is due to the nonoverlapping nature of the t$_{1u}$ and e$_u$ states.   For Co$_2$MnAl, the minority spin channel exhibits a gap, however, the t$_{1u}$ states  cross the Fermi level at the $\Gamma$ point indicating weak-half metallic properties. Similar behavior is shown by Co$_2$TiAl and Co$_2$VAl,  where the gap is present but the Fermi level lies within the conduction band. In Co$_2$TiAl, as we can see in the figure, the e$_u$ states just touch the Fermi level at the $\Gamma$ point and the bands originating from these states cross the Fermi level. However, in Co$_2$VAl, though we have a direct gap at the $\Gamma$ point, a point which reflects the characteristics of states in real space, as we move away from this point i.e. towards $X$ point, the bands originating from the e$_g$ states are lower in energy than the bands originating from the e$_u$ states and hence we will have an indirect gap between the $\Gamma$ and $X$ points. 

\begin{figure}[h]
	\centering
	\includegraphics[width=1\linewidth,height=0.25\textheight]{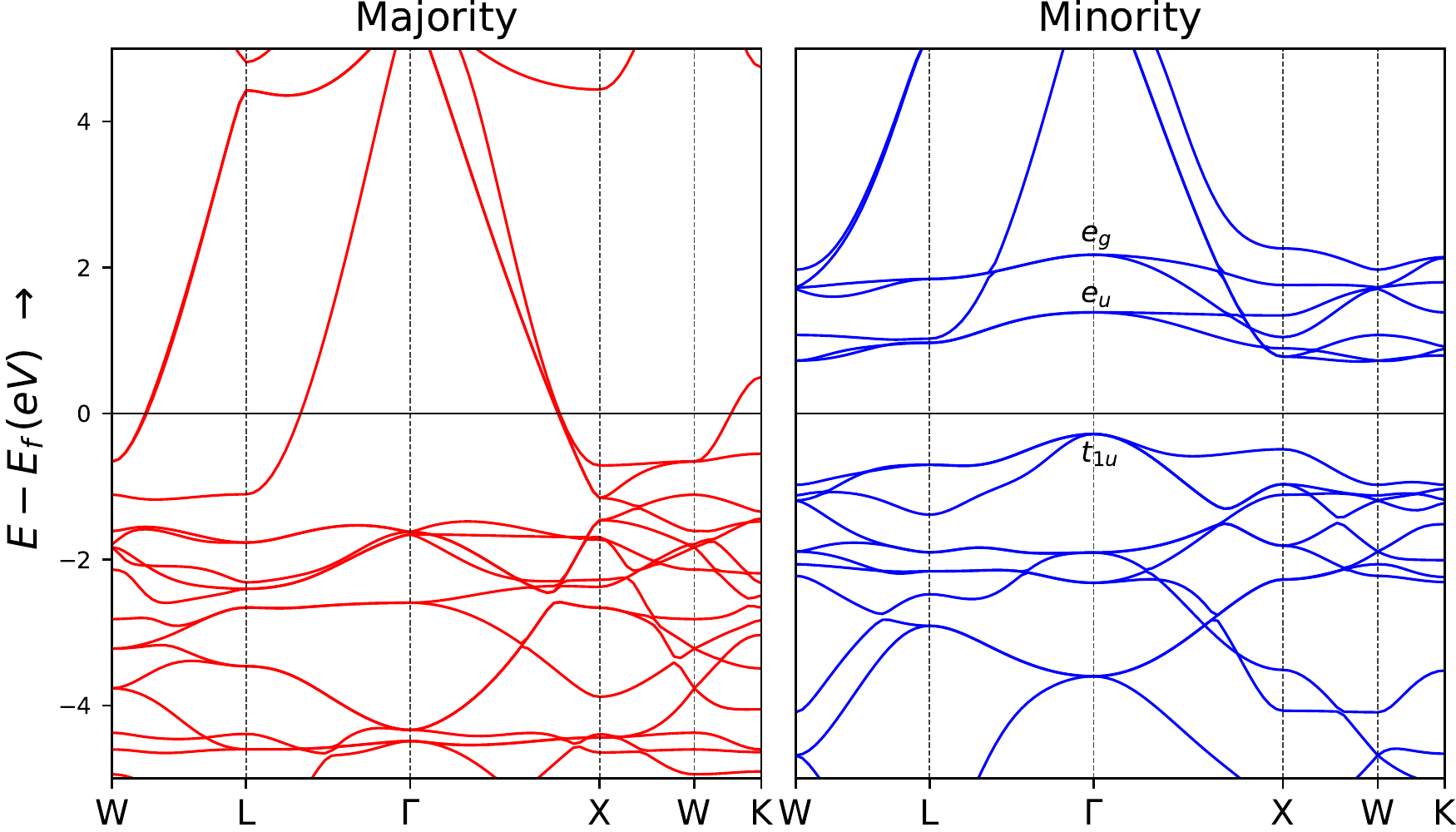}
	\caption{Calculated electronic band structure of the compound Co$_2$FeAl using the optimized lattice parameter and the GGA$ +U $ calculation.}
	\label{fig:band3}
\end{figure}

In the case of Co$_2$FeAl, depending on the lattice
constant, there is a very small gap or a narrow region of vanishing DOS.
This is due to the fact that either there is a tiny indirect $ \Gamma $-$X$ gap
or the valence band crosses the Fermi level at the $ \Gamma $-point while the
conduction band crosses the Fermi level at the $X$ point creating a
region of very vanishing DOS. The band structure using the optimized lattice parameter is shown in figure \ref{fig:band2}. However, a significant gap is observed after using GGA$+U$ as shown in figure \ref{fig:band3}. The explanation for the use of GGA$+U$ in the case of Co$_2$FeAl will be discussed separately later.

\par In the table \ref{tab:table3}, we present the individual and total spin magnetic moments of the Co$_2$ZAl compounds for the optimized lattice parameters. From the table, it is clear that both Co(A) and Co(C) have the same magnetic moment. This is due to the fact that both atoms, though they are at different sublattices, are chemically equivalent since they are surrounded by four Z atoms and four Al atoms. Here, Z  and Al atoms are surrounded by eight cobalt atoms as first neighbors, and hence cobalt atoms arrange themselves in an octahedral symmetry.

\par As we move from top to bottom, with the increase in the atomic number of the Z atoms, one can notice that the total spin magnetic moment of the compound increases roughly by $1\mu_B$. Only the compounds Co$_2$CrAl and Co$_2$FeAl strictly follow the Slater Pauling rule whereas other compounds deviate from it having non-integer values of the total magnetic moment. The integer value for Co$_2$FeAl is obtained only  when considering the on-site electronic correlations. It is also important to note that the magnetic moment of Co atoms increases slowly whereas the moment of Z atoms starting from negative value increases rapidly exceeding the value of Cobalt atoms for Z = Cr, Mn, and Fe. One can notice that the spin magnetic moment of the Al atom, in all cases, is negligibly small and anti-parallel with respect to the spin moments of the Co atoms. The magnetic moment of Al is induced by neighboring Co and Z atoms. This polarization of Al atoms increases with the increase in the magnetic activity of Z atoms  and the effect of this can be seen on the spin magnetic moment of the Al atoms.

\begin{table}[hp]
	\caption{Properties of compounds Co$ _2 $ZAl with Z = Sc, Ti, V, Cr, Mn and Fe calculated using optimized lattice parameter.}
	\label{tab:table3}
	\resizebox{0.48\textwidth}{!}{%
		\begin{tabular}{@{} l l c c c c c c c @{}}
			\toprule
			\multicolumn{1}{l}{\multirow{2}{*}{\begin{tabular}[l]{@{}l@{}}Compounds\\ (Co$_2$ZAl)\end{tabular}}} &
			\multicolumn{1}{l}{\multirow{2}{*}{\begin{tabular}[c]{@{}c@{}}a$_{exp}$\\ ($\AA{}$)\end{tabular}}} &
			\multirow{2}{*}{\begin{tabular}[l]{@{}c@{}}a$_{opt}$\\ ($\AA{}$)\end{tabular}} &
			\multirow{2}{*}{\begin{tabular}[l]{@{}c@{}}$\Delta E$\\ (eV)\end{tabular}} &
			\multicolumn{5}{c}{Magnetic moments ($\mu _b$)} \\ \cline{5-9} 
			\multicolumn{1}{c}{} &
			\multicolumn{1}{c}{} &
			&
			&
			\multicolumn{1}{l }{\begin{tabular}[c]{@{}l@{}}Total\\ {[}M$_T${]}\end{tabular}} &
			\multicolumn{1}{l }{\begin{tabular}[c]{@{}l@{}}Co{[}A{]}\\ {[}m$_A${]}\end{tabular}} &
			\multicolumn{1}{l }{\begin{tabular}[c]{@{}l@{}}Co{[}C{]}\\ {[}m$_C${]}\end{tabular}} &
			\multicolumn{1}{l }{\begin{tabular}[c]{@{}l@{}}Z{[}B{]}\\ {[}m$_B${]}\end{tabular}} &
			\multicolumn{1}{l }{\begin{tabular}[c]{@{}l@{}}Al{[}D{]}\\ {[}m$_D${]}\end{tabular}} \\ \toprule
			Co$_2$ScAl                                                     & \multicolumn{1}{c }{---} & 5.913 & ---   & 0.03 & 0.024 & 0.024 & -0.006 & -0.002 \\ 
			Co$_2$TiAl                                                     & 5.847$^{a,d}$            & 5.781 & 0.284 & 0.91 & 0.551 & 0.551 & -0.096 & -0.011 \\ 
			Co$_2$VAl                                                      & 5.722$^b$                & 5.686 & 0.251 & 1.98 & 0.926 & 0.925 & 0.217  & -0.020 \\ 
			Co$_2$CrAl                                                     & 5.727$^c$                & 5.640 & 0.503 & 3.00 & 0.847 & 0.847 & 1.379  & -0.047 \\
			Co$_2$MnAl                                                     & 5.749$^a$                & 5.632 & 0.517 & 4.01 & 0.824 & 0.824 & 2.504  & -0.064 \\
			\begin{tabular}[c]{@{}l@{}}Co$_2$FeAl\\  (DFT$+U$)\end{tabular} & 5.730$^{a,c}$            & 5.643 & 1.156 & 5.00 & 1.267 & 1.267 & 2.850  & -0.099 \\ 
			\begin{tabular}[c]{@{}l@{}}Co$_2$FeAl\\   (GGA)\end{tabular}     & 5.730$^{a,c}$            & 5.632 & ---   & 4.98 & 1.246 & 1.246 & 2.691  & -0.053 \\ \bottomrule
		\end{tabular}%
	}
	\\{\tiny Reference: $^a$ref\cite{BUSCHOW19831},$^b$ref\cite{Kandpalmainpaper}, $^c$ref\cite{BUSCHOW198190}, $^d$ref\cite{cta}}
\end{table}     

\par For the compounds with Z = V, Cr, and Mn, one can notice that the magnetic moment of Z atoms increases roughly by $1 \mu_B$ as we move downwards in the table. However, for the Fe atom, there is no such increase  and the magnetic moment of the Fe atom remains almost the same as of the Manganese atom. Also, the individual spin magnetic moments of the Co atoms scarcely changes when we replace Mn for Cr. The nature of the variation of the spin magnetic moments observed in our case are in good agreement with results in reference \cite{Galanakis2002} though the value of the magnetic moments are slightly different. Readers are referred to reference \cite{Galanakis2002} for the detailed explanation of these causes.   
   
\subsection{Effect of lattice parameter variation}  

\par In this section, the consequences of the lattice parameter variation on the electronic and magnetic properties of Co$_2$ZAl (Z = V, Cr, Mn, and Fe) will be discussed in detail. Table \ref{tab:tabl1} presents the variation  on the magnetic moments of the compounds Co$_2$ZAl (Z = V, Cr, Mn, Fe) with the deviation in lattice parameter from the experimental value. Although we have studied the properties for a larger range of lattice constants, here we will present  the data for the contraction and expansion of the lattice parameter by $\pm 2\%$.

\par
\begin{figure}[h]
	\centering	\includegraphics[width=1\linewidth]{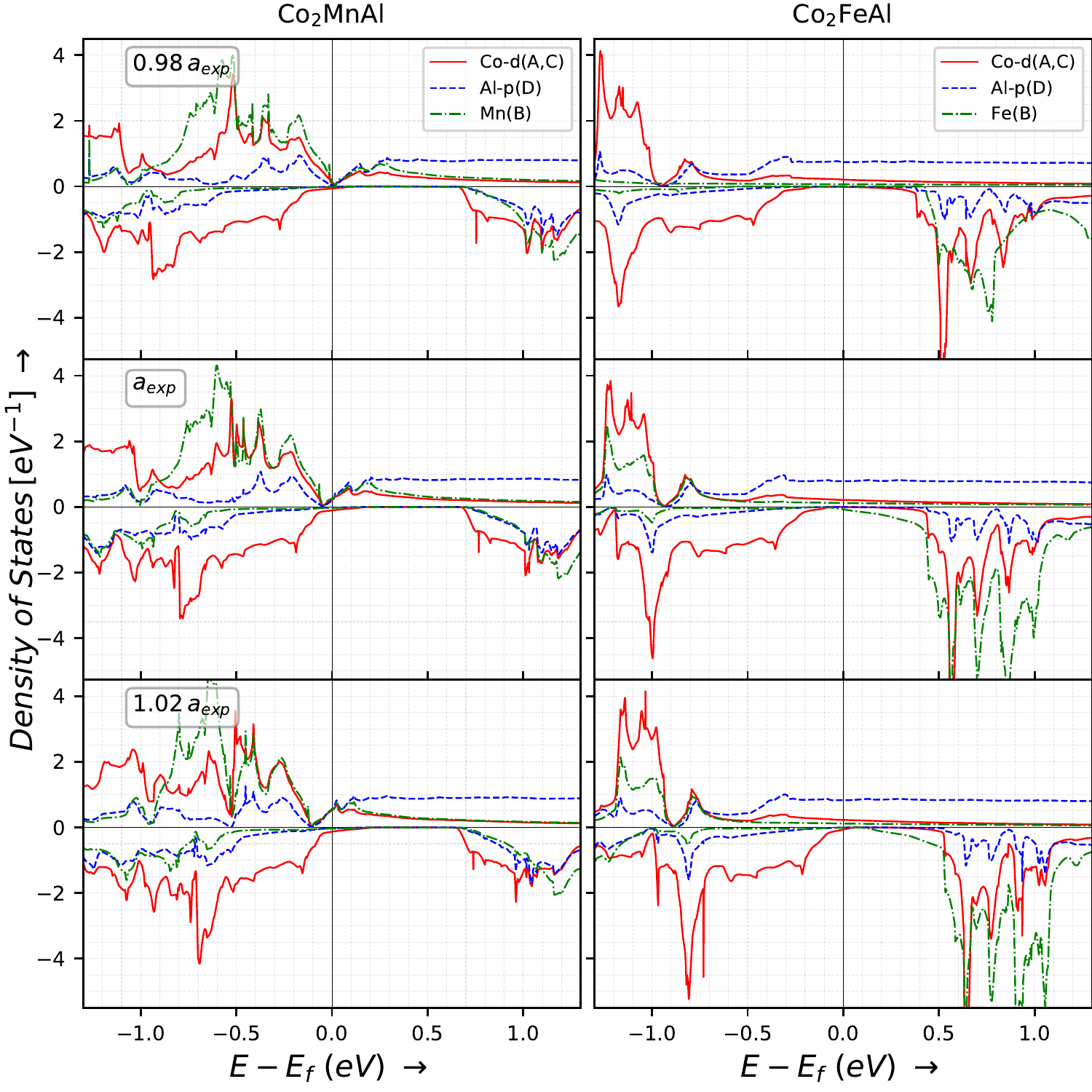}
	\caption{Partial DOS for the  compounds Co$_2$MnAl and Co$_2$FeAl. For Co$_2$FeAl, the PDOS obtained is when employing the GGA$+U$ scheme. Here, the negative DOS values correspond to the spin-down channel.\\
		(\textbf{Note:} DOS of Al-p states had been scaled ten times the original value for sake to clarity in visualization.)}
	\label{fig:03}
\end{figure}

\begin{figure}[h]
	\centering
	\includegraphics[width=1\linewidth]{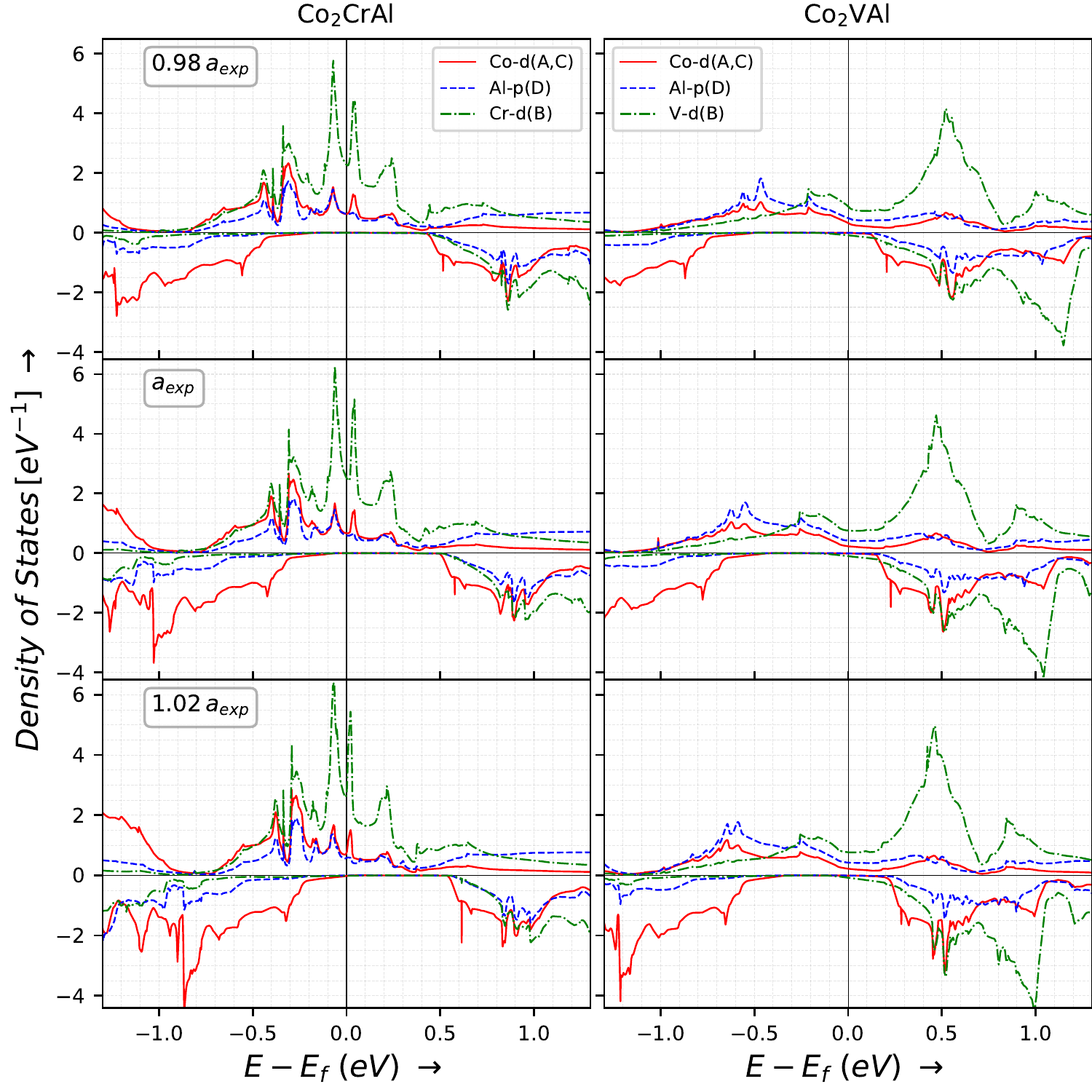}
	\caption{Partial DOS of compounds Co$_2$CrAl and Co$_2$VAl. Details as in figure \ref{fig:03}.}
	\label{fig:04}
\end{figure}

 It is interesting to analyze the behavior of the individual spin magnetic moments with the change in the lattice parametesr. We can remark that as we expand the lattice parameter the magnetic moment of the V atom decreases whereas for Cr, Mn, and Fe the opposite behavior can be noticed. However, in the case of Co atoms, the situation is more complicated and the spin magnetic moment does not follow a particular pattern. One should note that the change in the spin magnetic moment is not easy to explain in all cases. As we expand the lattice parameter, two phenomena are accompanied simultaneously. The atomic distance increases due to which the hybridization between \textit{d} orbitals of neighboring atoms decreases (see figure \ref{fig:03} and \ref{fig:04}) and individual atoms in lattice crystal manifest more atomic behavior. This is the reason behind the increase in the spin magnetic moment of Cr, Mn, and Fe atoms with the increase in lattice parameter; the decrease in magnetic moment of V is obvious as it is not magnetic itself. For Cobalt atoms, the condition is more complex. On one hand, they tend to increase the spin magnetic moment with increase in lattice constant like other magnetic atoms manifesting more atom-like behavior. This behavior can be seen in the case of Co$_2$VAl, where hybridization between Co and V is very weak compared to the other three compounds. On the other hand, the larger spin magnetic moment of Cobalt atoms on C site means the larger exchange splitting between occupied and unoccupied \textit{d} states. This interaction is not always easy to perceive and there are cases like Co$_2$CrAl where the spin magnetic moment of Co atoms decreases and there is an instance where it increases as in Co$_2$FeAl and circumstances like Co$_2$MnAl where the spin magnetic moment is maximum for experimental lattice parameter.

\par However, it is evident from table \ref{tab:tabl1} that the variation on the lattice parameter, within a short-range, has little or no effect on the total spin magnetic moment of the compounds although the individual spin magnetic moments vary significantly. The reason behind the stability of the total magnetic moment lies in the fact that such variation on one lattice site is counterbalanced by another variation of similar magnitude on another lattice site. 

\par In figure \ref{fig:03} and \ref{fig:04} we  present the atom and spin-resolved DOS for the four compounds under study. Here, one can clearly notice an increase in bandwidth with the decrease in lattice constant and vice versa. As we decrease the lattice constant, the Fermi-level shifts higher in energy closer to the conduction states whereas increase in lattice parameter pushes the Fermi-level towards the occupied states, lower in energy. Previously, such behavior has been noticed in half-Heusler alloys like NiMnSb and CoMnSb, and attempts  have been made to understand this  effect\cite{halfheusler}. 
This phenomenon can be understood by taking into account the nature of the \textit{p} electrons. As we know, \textit{p} electrons in Al atoms are well scattered and lie deep below the Fermi level; the contraction of the lattice parameter pushes the lower-lying delocalized \textit{p} electrons higher in energy, which increases the hybridization with localized \textit{d} electrons, resulting in  the shift of the Fermi level higher in energy. When we increase the lattice parameter the reverse effect can be seen(see figure\ref{fig:03} and \ref{fig:04}).

\begin{table}[h]
	\caption{Properties of the compounds Co$ _2 $ZAl with Z = V, Cr, Mn and Fe calculated for $\pm 2\%$ change in the experimental lattice parameter.}
	\label{tab:tabl1}
	\resizebox{0.48\textwidth}{!}{%
		\begin{tabular}{@{} l l c c c c c c c @{} }
			\toprule
			\multicolumn{2}{ l }{\multirow{2}{*}{\begin{tabular}[c]{@{}l@{}}Compounds\\ (Co$_2$ZAl)\end{tabular}}} &
			\multirow{2}{*}{a ($\AA{}$)} &
			\multirow{2}{*}{\begin{tabular}[c]{@{}c@{}}$\Delta E$\\ (eV)\end{tabular}} &
			\multicolumn{5}{c }{Magnetic moments ($\mu _b$)} \\ \cline{5-9} 
			\multicolumn{2}{ l }{} &
			&
			&
			\multicolumn{1}{l }{\begin{tabular}[c]{@{}l@{}}Total\\ {[}M$_T${]}\end{tabular}} &
			\multicolumn{1}{l }{\begin{tabular}[c]{@{}l@{}}Co{[}A{]}\\ {[}m$_A${]}\end{tabular}} &
			\multicolumn{1}{l }{\begin{tabular}[c]{@{}l@{}}Co{[}C{]}\\ {[}m$_C${]}\end{tabular}} &
			\multicolumn{1}{l }{\begin{tabular}[c]{@{}l@{}}Z{[}B{]}\\ {[}m$_B${]}\end{tabular}} &
			\multicolumn{1}{l }{\begin{tabular}[c]{@{}l@{}}Al{[}D{]}\\ {[}m$_D${]}\end{tabular}} \\ \toprule
			\multirow{3}{*}{Co$_2$VAl} &
			$a_{-2\%}$ &
			5.608 &
			0.251 &
			1.99 &
			0.904 &
			0.905 &
			0.245 &
			-0.018 \\ 
			&
			$a_{exp}$ &
			5.722 &
			0.196 &
			1.99 &
			0.953 &
			0.953 &
			0.182 &
			-0.021 \\ 
			&
			$a_{+2\%}$ &
			5.836 &
			0.163 &
			2.00 &
			0.998 &
			0.998 &
			0.130 &
			-0.024 \\ \hline
			\multirow{3}{*}{Co$_2$CrAl} &
			$a_{-2\%}$ &
			5.613 &
			0.503 &
			3.00 &
			0.855 &
			0.856 &
			1.355 &
			-0.046 \\ 
			&
			$a_{exp}$ &
			5.727 &
			0.522 &
			3.00 &
			0.850 &
			0.851 &
			1.396 &
			-0.052 \\ 
			&
			$a_{+2\%}$ &
			5.842 &
			0.479 &
			3.01 &
			0.837 &
			0.834 &
			1.466 &
			-0.060 \\ \hline
			\multirow{3}{*}{Co$_2$MnAl} &
			$a_{-2\%}$ &
			5.634 &
			0.517 &
			4.01 &
			0.826 &
			0.826 &
			2.505 &
			-0.064 \\ 
			&
			$a_{exp}$ &
			5.749 &
			0.460 &
			4.12 &
			0.830 &
			0.830 &
			2.633 &
			-0.072 \\ 
			&
			$a_{+2\%}$ &
			5.864 &
			0.402 &
			4.12 &
			0.787 &
			0.788 &
			2.751 &
			-0.079 \\ \hline
			\multirow{3}{*}{\begin{tabular}[c]{@{}l@{}}Co$_2$FeAl\\ (DFT$ +U $)\end{tabular}} &
			$a_{-2\%}$ &
			5.615 &
			1.156 &
			5.00 &
			1.265 &
			1.265 &
			2.838 &
			-0.095 \\ 
			&
			$a_{exp}$ &
			5.730 &
			1.016 &
			5.00 &
			1.276 &
			1.276 &
			2.878 &
			-0.109 \\ 
			&
			$a_{+2\%}$ &
			5.845 &
			0.596 &
			5.00 &
			1.291 &
			1.291 &
			2.911 &
			-0.123 \\ \hline
			\multirow{3}{*}{\begin{tabular}[c]{@{}l@{}}Co$_2$FeAl\\ (GGA)\end{tabular}} &
			\multicolumn{1}{c }{$a_{-2\%}$} &
			5.615 &
			--- &
			4.98 &
			1.246 &
			1.246 &
			2.684 &
			-0.052 \\ 
			&
			\multicolumn{1}{c }{$a_{exp}$} &
			5.730 &
			--- &
			4.99 &
			1.248 &
			1.248 &
			2.7333 &
			-0.061 \\ 
			&
			\multicolumn{1}{c }{$a_{+2\%}$} &
			5.845 &
			0.053 &
			5.01 &
			1.254 &
			1.254 &
			2.789 &
			-0.070 \\ \bottomrule
		\end{tabular}%
	}
\end{table}

\subsection{Effect of correlation on electronic structure}

For the precise modeling of the electronic structure and property of different classes of materials, especially semiconductors, GGA$ +U $ or LDA$ +U $ is one of the cheap alternative tools to GW calculations and they are  widely used nowadays. GGA(LDA) underestimates the gap in semiconductors and the inclusion of the $ U $ opens the gap by including the electrostatic repulsion term between the electrons of an atom residing in the same shell. However, one should be cautious using DFT+$U$ in metals and magnets as its use is not always justified and it should be used only when the electrons are strongly correlated. Keeping this in mind, DFT$+U $ calculations were performed with the GGA parametrization for the compounds Co$_2$ZAl  [Z = Cr, Mn, and Fe]. The values of the $ U $ and $ J $ parameters used in calculation are tabulated in table \ref{tab:hubbardU}. 

\par The properties calculated are presented in table \ref{tab:dftuprop}. When comparing the GGA and GGA$+U$ calculations at the experimental lattice parameter(see tables \ref{tab:tabl1} and \ref{tab:dftuprop}), we find that the inclusion of the electronic correlations does not affect the total magnetic moments except for  Co$_2$MnAl in which the Fermi level shifts deep into the valence band and the total spin magnetic moment deviates largely from its experimental value $ 4.04\,\mu_B $\cite{BUSCHOW198190}. This unrealistic deviation of the magnetic moment of Co$_2$MnAl, in our case, is in good agreement with the previous literature\cite{dftueffect}, except that their deviation is far larger in comparison to our calculation. One can see from table \ref{tab:tabl1} and \ref{tab:dftuprop} that the increase in the spin magnetic moments of the compound is due to the increase in the individual spin magnetic moments of the Mn atoms. However, such a sharp increase is not noticed in the case of the Cr and Fe atoms. Though there is no  change in the total spin magnetic moment for Co$_2$CrAl, the half-metallic gap, which already exists in $ GGA $, is further widened by the inclusion of $ U $ in the calculation.

\par In the case of Co$_2$FeAl, our calculation shows that a very small gap exists within GGA  but a semiconducting behavior is not observed since the Fermi level crosses the conduction band. However, the inclusion of electronic correlations employing the GGA$+U$ scheme  induces a perfect half-metallic gap with an integer total spin magnetic moment of 5.0 $ \mu_B $ which agrees with the Slater-Pauling rule\cite{Galanakis2002}. In an attempt to study this effect meticulously, we have tried to inspect the changes in PDOS and total DOS of Co$_2$FeAl around the Fermi level with and without the inclusion of $ U $. 
\begin{figure}[h]
	\centering	\includegraphics[width=1\linewidth]{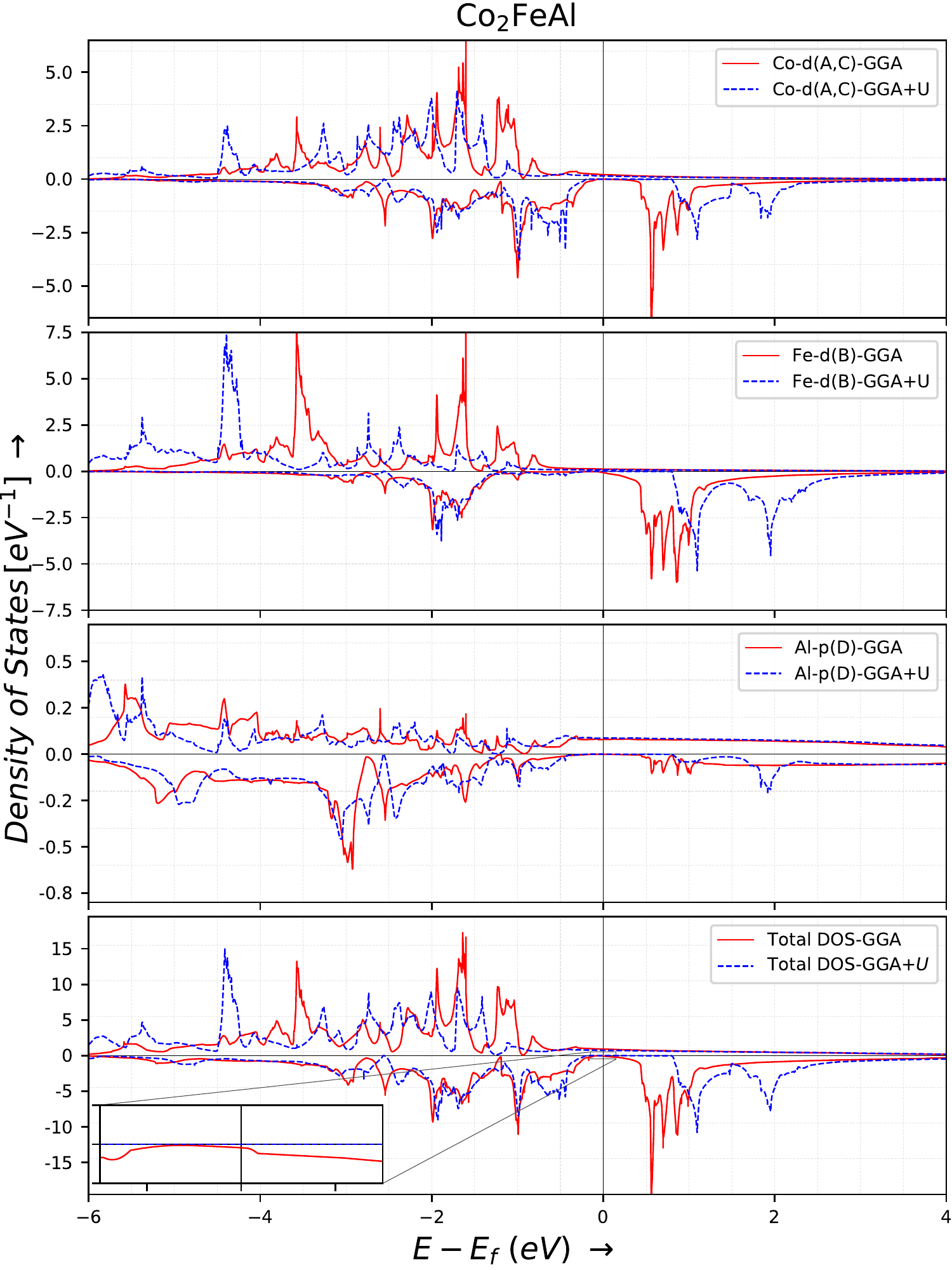}
	\caption{Partial DOS and total DOS for Co$_2$FeAl using both GGA and  GGA$+U$ at the  experimental lattice parameter.\label{fig:05}}
\end{figure}

It is clear from figure \ref{fig:05} that the loss in half-metallicity in the compound within the standard GGA  calculation is due to the presence of states at the Fermi level originating from the Fe atoms. Though the gap is smaller in the PDOS of the Co atoms with respect to the Fe atoms, the Fermi level lies within the gap in their case and hence it is not responsible for the destruction of half-metallicity in Co$_2$FeAl.

\par The  GGA$+U $  scheme in conjunction with the FLL functional for the double counting term induces the half-metallicity in Co$_2$FeAl. As we can see in figure \ref{fig:05},  the inclusion of the correlations increases the exchange splitting between the occupied majority-spin bands (which are shifted lower in energy) and the unoccupied minority-spin bands (which are shifted higher in energy). Moreover, all bands become more narrow. These changes influence also the occupied minority-spin bands. If we focus in the region just below the Fermi level, the smaller weight of the occupied majority-spin states and the narrowing of the occupied minority-spin bands lead to vanishing occupied minority-spin states just below the Fermi level resulting to the desired half-metallicity.  

\par It is important to note that the use of GGA$ +U $ in the case of Co$_2$CrAl and Co$_2$MnAl is not justified as in former the gap largely deviates from the bandgap obtained for the  experimental lattice parameter \cite{Co2CrAl} whereas in the latter GGA$+U$ produces an unrealistic picture of spin magnetic moments. As can be seen in table  \ref{tab:dftuprop} the total spin magnetic moment of Co$_2$MnAl is about 4.86 $\mu_B$ deviating considerably from the value of about 4 $\mu_B$ obtained by simple GGA as shown in table \ref{tab:table3} and as expected from the Slater-Pauling rule for half-metals. Also, the valence electrons of these compounds are less i.e 27 and 28 respectively compared to Co$_2$FeAl which has 29 valence electrons. Literature has shown that the use of DFT$ +U $ schemes like GGA$+U$ is necessary to explain the magnetic moments in Co$_2$FeSi\cite{KandpalcompareMnFe}, which has 30 valence electrons, however, for Co$_2$MnSi which has 29 valence electrons like Co$_2$FeAl, its use is not justified. Thus the effect of GGA$+U$ does not  depend only on the total number of valence electrons but also on the specific chemical formula of the compound under study. 
\begin{table}[h]
	\caption{Properties of Co$_2$ZAl [Z = Cr, Mn and Fe] calculated using GGA$ +U $ at the experimental lattice parameter.}
	\label{tab:dftuprop}
	\resizebox{0.48\textwidth}{!}{
	\begin{tabular}{lccccc}
		\toprule
		\multicolumn{1}{c}{\multirow{2}{*}{\begin{tabular}[c]{@{}c@{}}Compounds\\ (Co$_2$\,Z\,Al)\end{tabular}}} & \multicolumn{5}{c}{Magnetic moments ($\mu _b$)} \\ \cline{2-6} 
		\multicolumn{1}{c}{} &
		\multicolumn{1}{l}{\begin{tabular}[c]{@{}l@{}}Total\\ {[}M$_T${]}\end{tabular}} &
		\multicolumn{1}{l}{\begin{tabular}[c]{@{}l@{}}Co{[}A{]}\\ {[}m$_A${]}\end{tabular}} &
		\multicolumn{1}{l}{\begin{tabular}[c]{@{}l@{}}Co{[}C{]}\\ {[}m$_C${]}\end{tabular}} &
		\multicolumn{1}{l}{\begin{tabular}[c]{@{}l@{}}Z{[}B{]}\\ {[}m$_B${]}\end{tabular}} &
		\multicolumn{1}{l}{\begin{tabular}[c]{@{}l@{}}Al{[}D{]}\\ {[}m$_D${]}\end{tabular}} \\ \toprule
		Co$_2$CrAl                                                                                               & 3.00   & 0.910   & 0.911   & 1.503   & -0.102   \\ 
		Co$_2$MnAl                                                                                               & 4.86   & 0.866   & 0.865   & 3.425   & -0.117   \\ 
		Co$_2$FeAl                                                                                               & 5.00   & 1.276   & 1.276   & 2.878   & -0.109   \\ \bottomrule
	\end{tabular}%
	}
\end{table}

For Co$_2$FeAl, which has the same valence electrons as Co$ _2 $MnSi, one can not draw a definitive conclusion whether the use of GGA$ +U $ is justified or not since there is no known experiment on a bulk crystal of Co$_2$FeAl. The change in the spin magnetic moments, formation, and cohesive energies with respect to standard GGA is not significant but the change in the gap is important.  In the case of Co$_2$FeAl, since it has one more valence electron in the 3\textit{d} site rather than the \textit{sp} site in comparison to Co$ _2 $MnSi, one could expect this to be the limiting case where the on-site electronic correlations might be important.  

\section{Conclusion}\label{sec4}

Exploiting a pseudopotential \textit{ab-initio} electronic band structure method we have investigated the structural, electronic, and magnetic properties of Co$_2$ZAl compounds where Z is a transition metal atom from Sc to Fe. After verifying the stability of the compounds using formation and cohesive energies calculations, we have confirmed the half-metallic (or almost half-metallic) character of all compounds with the exception of Co$_2$FeAl. We have studied the effect of the lattice parameter variation on the electronic and magnetic properties of the compounds and have shown that  small variations do not affect significantly the total spin magnetic moment and the half-metallic property.  We have employed the GGA$+U$ scheme to account for the electronic correlations. Our results show that the effect of GGA$+U$ is materials specific. There are cases like Co$_2$MnAl where the inclusion of $U$ leads to unrealistic results. In the case of  Co$_2$FeAl its use might be important to describe electronic structure since it leads to half-metallicity contrary to standard GGA. Detailed experimental data are needed to reach a safe conclusion on the suitability of GGA$+U$ to describe the properties of Co$_2$FeAl. 

We expect our results to further strengthen the interest and the research on the Co-based Heusler compounds in order to be implemented in realistic spintronic devices.              
   
\bibliography{mybibfile}

\end{document}